\documentclass[11pt,twoside]{acta}
\usepackage{epsfig,float,floatflt,graphics,amsmath,amssymb,pstricks,wasysym,
stmaryrd,enumerate,epic,eepic,rotating}
\SetPages{000}{000}
\SetVol{00}{2005}

\begin{document}

\begin{Titlepage}
  \Title{Seismic diagnostics of mixing beyond the convective core in intermediate mass
main-sequence stars}
  \Author{ B.~L.~P~o~p~i~e~l~s~k~i~$^1$ \and
  	   W.~A.~D~z~i~e~m~b~o~w~s~k~i~$^{1,2}$}
{$^1$ Warsaw University Observatory, Al.Ujazdowskie~4,~00-478~Warsaw, Poland\\
 $^2$ Nicolaus Copernicus Astronomical Center, ul.Bartycka~18, 00-716~Warsaw, Poland\\
e-mail: blapo@fuw.edu.pl, wd@astrouw.edu.pl}

\end{Titlepage}

\Abstract{We study prospects for seismic sounding the layer of a partial
mixing above the convective core in main-sequence stars with
masses in the 1.2 -- 1.9 M$_\odot$ range. There is an initial
tendency to an increase of convective core mass in such stars and
this leads to ambiguities in modeling. Solar-like oscillations
are expected to be excited in such objects. Frequencies of
such oscillations provide diagnostics, which are sensitive to the 
structure of the innermost part of the star and they are known as 
the small separations. We construct evolutionary models of stars in this mass
range assuming various scenarios for element mixing, which includes
formation of element abundance jumps, as well as semiconvective
and overshooting layers. We find that the three point small
separations employing frequencies of radial and dipole modes
provide the best probe of the element distribution above the
convective core. With expected accuracy of frequency
measurement from the space experiments, a discrimination between
various scenarios should be possible.
}
{stars: variable, stars: oscillations, stars: solar-type, stars: convection}

\section{Introduction}
\label{s:1}

The extent of mixing beyond the convective core has been debated
for several decades and the issue remains
unsettled. For stars with masses above 1.9 M$_\odot$ in the main-sequence evolutionary phase
the convective core shrinks in mass and the issue concerns the extent of the overshooting
beyond the boundary determined by Schwarzschild criterion (1906), which in this case is
identical to Ledoux criterion (1947). At lower masses there is a tendency for the convective
core expansion. Outside of the expanding core, in the environment more abundant in
hydrogen, the radiative temperature gradient must be higher than below.
Consequently, according to Schwarzschild criterion, such layer must be unstable.
It is known (e.g. Kato, 1966) that in the latter case the instability takes form
of growing oscillatory mode. The nonlinear development of the motion in the layers
with preexistent chemical inhomogeneities is particularly difficult to study.

This problem has been first encountered by Schwarzschild~and H\"arm (1958) at very
high stellar masses, above 30 M$_{\odot}$. In the case of these very massive stars
the growth of the core
was caused by the temperature increase in time, hence lowering $\beta = p_{gas} / p$,
hence lowering adiabatic gradient, $\nabla_{ad}.$
To overcome the problem these authors introduced the concept of semiconvective zone outside
the convective core, where hydrogen abundance rises steadily in such a way
that convective neutrality is preserved. They presumed that in this layer
there is a slow convection which carries negligible amount of energy, but
sufficient for the required partial mixing.

In the stars with masses lower than 1.9~M$_\odot,$ the tendency for core expansion is caused
by an increase of relative contribution of CNO cycle to
energy production, which results in an increase of radiative gradient.
This was first observed by Crowe~and Mitalas (1982), who followed
Schwarzschild~and H\"arm (1958) recipe.
A physical support for the semiconvective model came from Stevenson (1979), who
argued that occasional wave breaking enforces the neutral stability according to
Schwarzschild criterion. This argument was subsequently questioned by Spruit
(1992), who inspired by simple terrestrial experiment, argued that a thin
layer with very steep rise of hydrogen abundance gives rise to an entropy barrier
force preventing element mixing.
Above this layer there would be another convective
zone. Numerical simulations of Biello (2001) provided certain support for
Stevenson's picture. However, we should stress that in his simulations the
adopted Prandtl number was substantially higher than in stars.
These two scenarios do not exhaust all the possibilities. There is no reason
to assume that overshooting stops entirely as soon as the tendency to core expansion appears.
Overshooting may erase the hydrogen abundance jump above the expanding core.
For this reason in addition we will also consider scenarios with adjustable overshooting
distance.

We believe that the best prospect for distinguishing between various scenarios
may come from analysis of solar-type oscillation frequencies. In the
mass range 0.9 -- 2 M$_{\odot}$ we expect
excitation of high order acoustic oscillations (p-modes) by the same mechanism
that excites solar oscillation (Houdek~\etal 1999). In fact the prediction is that
the intrinsic oscillation amplitudes in these more massive stars would be
higher than those in the Sun.

We will show in this work that with the accuracy of frequency measurements
expected from dedicated space missions like COROT (Baglin~\etal 2002),
MOST (Walker~\etal 1998) or EDDINGTON (Roxburgh~and Favata 2003) we will be able
to distinguish various models of element mixing in deep stellar interior.

In Section~\ref{s:2} we describe three scenarios for mixing above the
expanding core. There we also give a short overview of our modeling procedures.
In Section~\ref{s:3} we introduce two types of small separations.
The small separations are further used to examine the effect of mixing in a grid of models
across the domain of solar-like stars (Section~\ref{s:41}). The possibility of distinguishing
chemical profiles through small separations is discussed in the next section (Section~\ref{s:42}).
We summarize our work in Section~\ref{s:5}

\section{Alternative models of element mixing above the expanding core}
\label{s:2}

\begin{figure}
\centering
   \epsfig{figure=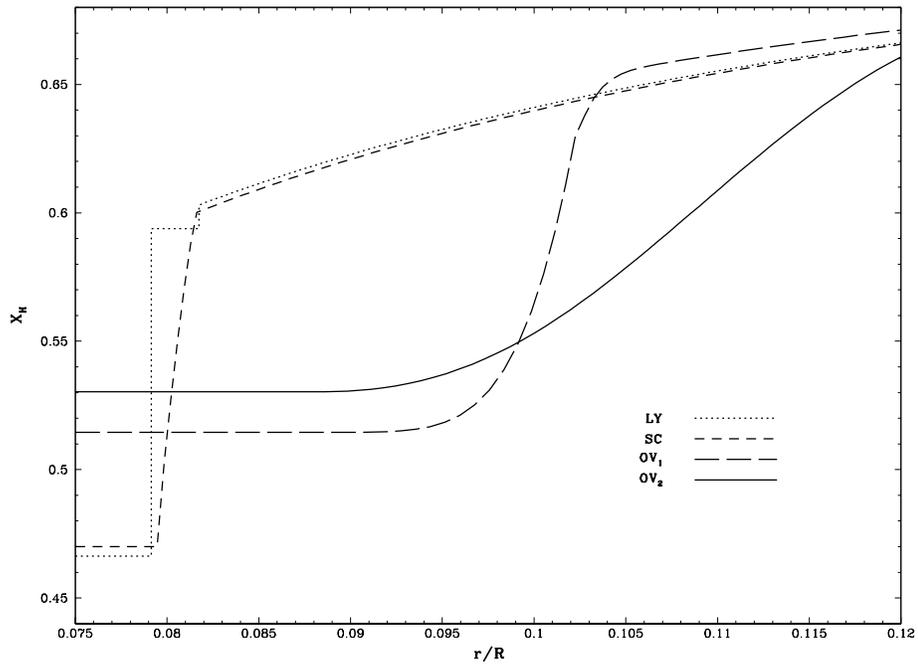,height=1.03 \linewidth,angle=-90}
\caption{Hydrogen abundance profile in the layer above the convective zone for a set
of slightly evolved 1.5 M$_\odot$ models.
Four curves refer to four mixing scenarios: layered, {\tt LY}, semiconvective, {\tt SC},
and two types of overshooting, {\tt OV$_1$} and
{\tt OV$_2$}, where $\alpha_{ov}=0.2$ and $0.5$ respectively.
The four models have the same radius of $1.68$ R$_\odot$. 
}
\label{f:1}
\end{figure}

In our work we consider three mixing scenarios in evolutionary models. Going from
minimum to maximum extent of element mixing, we examine:
\begin{enumerate}
  \item the layered model ({\tt LY}),
  \item the semiconvective model ({\tt SC}),
  \item the overshooting model ({\tt OV}).
\end{enumerate}
The {\tt LY} model is based on Spruit (1992) scenario. We approximate the two steep
hydrogen abundance rises by two composition jumps at the edges of the convective regions,
as seen in Figure~\ref{f:1}.
The discontinuities of density are explicitly taken into account in calculation of our
stellar models and their oscillation properties. The {\tt SC} model has both density and
temperature gradient continuous. However, the second derivatives of density and temperature
are discontinuous at both edges of the semiconvective zone.
In the {\tt OV} model, the extent of the overshooting above the convective core
is determined by the product $\alpha_{ov} H_p$, where $\alpha_{ov}$ is an adjustable
dimensionless parameter and $ H_p $ is the pressure distance scale calculated
at the edge of the core. Here we adopt smooth abundance profile,
leading to continuous density and its derivative. The profile of hydrogen
abundance in the overshooting zone, $ \left[ r_0,r_1=r_0+\alpha_{ov} H_p\right] $,
is given by $ X( r ) = X( r_0 ) + y(z),$ where
\begin{align}
 y(z)&=z^s \left[ \left( X'-s \Delta X\right) z - X' + (s+1) \Delta X \right],\\
 z&=\frac{r-r_0}{r_1-r_0}, \qquad \Delta X=X(r_1)-X(r_0),
 \qquad X'=(r_1-r_0) \left.\frac{d X}{d r}\right\vert_{r_1}
\end{align}
and $ s>1 $ is adjustable shape parameter.
The standard overshooting model is represented by $s\rightarrow\infty.$
In Figure~\ref{f:1} we show hydrogen abundance profile
obtained with different scenarios.

The influence of rotation on mixing is not necessarily negligible, however
its detailed discussion is beyond the scope of our paper.

In our code, which is an updated version of Paczy{\'n}ski's stellar evolution code
(1970)\footnote{Most of updates were made by M.Koz{\l}owski and R.Sienkiewicz.},
the chemical evolution is computed in an implicit way.
This requires additional routine to iterate
element abundances but is quite important to accurately follow the boundaries of
convective and semiconvective zones.
The chemical iteration scheme is not new. It was used e.g. by Crowe
and Mitalas (1982), VandenBerg (1983) and Shaller~\etal (1992).

The code includes:
\begin{itemize}
  \item the radiative opacities from OPAL project (Iglesias~and Rogers, 1996),
  the molecular opacities from Alexander~and Ferguson (1994),

  \item the equation of state EOS2001 from OPAL project (Rogers, 2001),

  \item the heavy element distribution from Grevesse~and Noels (1993).
\end{itemize}

The oscillation properties of models were calculated using a standard linear pulsation code
(Dziembowski, 1977), where the discontinuity was explicitly taken into account.

\section{Small separations}
\label{s:3}

The layers that we want to probe are located very close to the center. Low degree
high order p-modes reach down-there, however the question is whether there is enough
sensitivity to changes in this region to expect measurable effects.
The frequency combinations known to be particularly sensitive to stellar interior
are familiar small separations, see Christensen-Dalsgaard (1988), defined as
\begin{equation}
  d_{\ell,\ell+2}(n) = \nu_{n,\ell} - \nu_{n-1,\ell+2},
  \label{eq:ss02}
\end{equation}
for $\ell=0$ and $1$. 
Instead of $d_{13}$ it is better to use the three point separations, $d_{01},$ as
Roxburgh (1993) proposes,
\begin{equation}
  d_{01}(n) = \frac{\nu_{n,1} - 2\nu_{n+1,0} + \nu_{n+1,1}}{2}.
  \label{eq:ss01}
\end{equation}
because at $\ell=3$ we expect a considerable amplitude reduction.

First analytical expressions for small separations were derived by Tassoul (1980) from a
simple dispersion relation. However, her approximation is not satisfactory at quantitative
level, as pointed by Roxburgh and Vorontsov (1994), who derived improved expressions.
The price is that their complicated formula is not revealing. Therefore in our work
we rely solely on numerical calculations of eigenfrequencies.

\section{Discriminating between different mixing scenarios with small separations}
\label{s:4}

\subsection{Frequency dependence of small separations}
\label{s:41}

\begin{figure}
\centering
   \epsfig{figure=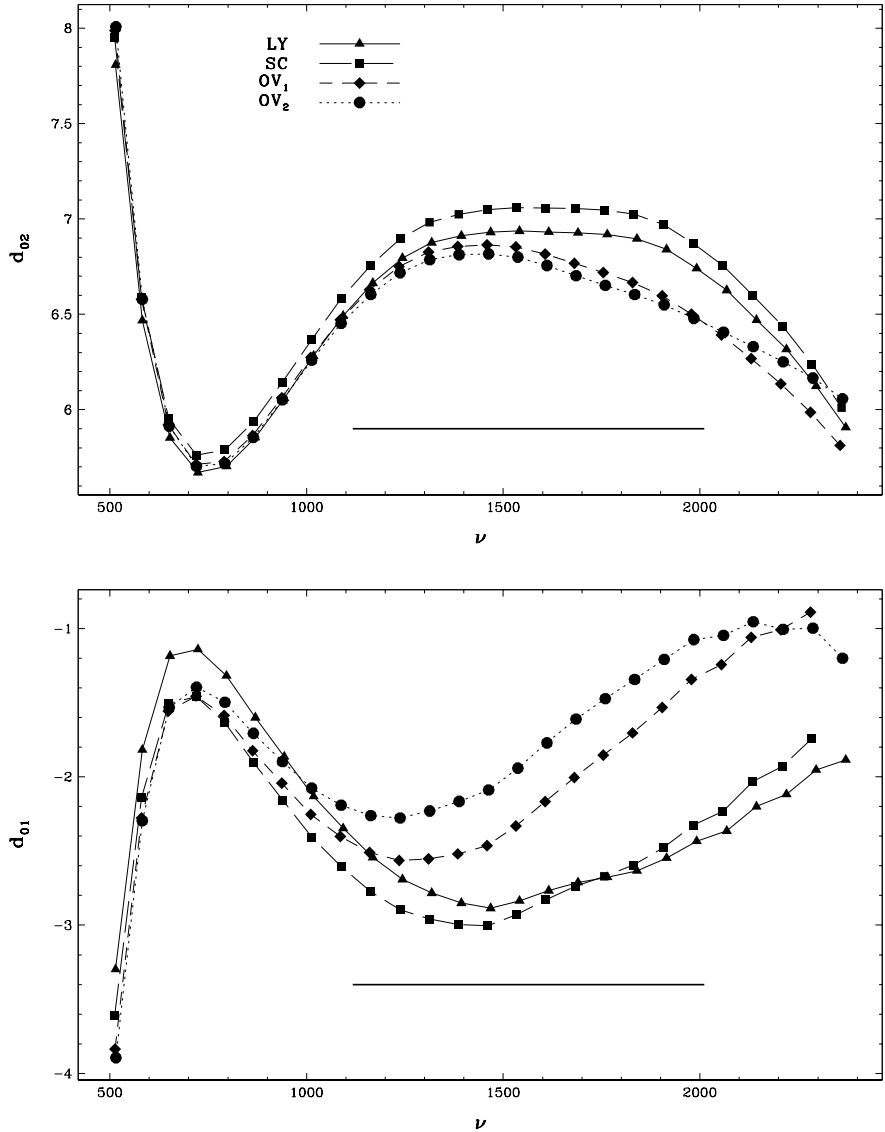,width=\linewidth}
\caption{Small separations for the four models from Figure~\ref{f:1},
layered, {\tt LY}, semiconvective, {\tt SC}, and two 
extents of overshooting, {\tt OV$_1$} and {\tt OV$_2$}.
The mean large frequency separations for the models are equal, $\overline{D}_0=74.5 \mu$Hz.
Thick horizontal line encompasses the n-range, defined in the text.
We see that there is more sensitivity to mixing scenarios in $d_{01}$ (lower panel).}
\label{f:2}
\end{figure}

The frequency dependence of small separations is not monotonic.
Figure~\ref{f:2} shows typical examples of $d_{02}$ and $d_{01}.$
Regardless of the mixing scenario, there is an overall similarity of the $\nu$-dependence
shapes. There is always a minimum of $d_{02}$ at $\nu=750 \mu$Hz and a corresponding
maximum of $d_{01}$ at the same frequency. The $d_{02}$ maximum at $\nu=1500 \mu$Hz
corresponds to minimum of $d_{01}.$

The upper limit of considered frequencies is set by 
$\nu_{\rm cutoff},$ which may be regarded as the highest frequency of the modes
we may ever expect to be seen in distant stars.
Although the peaks in the solar oscillation spectrum are seen above this frequency,
they are low and broad. The lower limit is set arbitrarily by an approximate validity
of asymptotics. However, the frequency that we may expect to measure in the near future
belongs to a significantly narrower range.
The horizontal line encompasses the range of mode orders, $n=15\div27,$ we are most
likely to detect. In determining this range we were guided by currently available
oscillation spectra for distant stars and by the frequency range of low degree solar
oscillation modes usable in helioseismic sounding. The summary of the observational
information relevant for determining this range is given in Table~\ref{t:solar-like}.

\begin{table}
\centering
\begin{tabular}{|c|ccc|r|ccc|l|}
 \hline
  object & \multicolumn{3}{c|}{frequency range $[\mu Hz]$} & $\overline{D}_0$ &
           \multicolumn{3}{c|}{$n$} & references\\
  \hline
  $\alpha$ Cen A & $1675$ & $\div$ & $3055$ & $106.2$ & $16$ & $\div$ & $29$ & Bedding~\etal (2004) \\
  $\eta$ Boo     & $600$ & $\div$ & $1100$ & $40.5$ & $15$ & $\div$ & $27$ & Kjeldsen~\etal (2003) \\
  Procyon A      & $300$ & $\div$ & $1400$ & $53.6$ & $6$ & $\div$ & $26$ & Marti{\'c}~\etal (2004) \\
  \hline
  Sun            & $2400$ & $\div$ & $3700$ & $135.5$ & $8$ & $\div$ & $33$ & Gabriel~\etal (1997) \\
  \hline
\end{tabular}
\caption{Ranges of frequency data for solar-like stars.
Corresponding radial orders, $n,$ are derived from the mean large frequency separations,
$\overline{D}_0.$ The solar data refers to low spherical degree modes whose frequencies are
accurately determined.}
\label{t:solar-like}
\end{table}

\subsection{The survey}
\label{s:42}

\begin{figure}
\centering
   \epsfig{figure=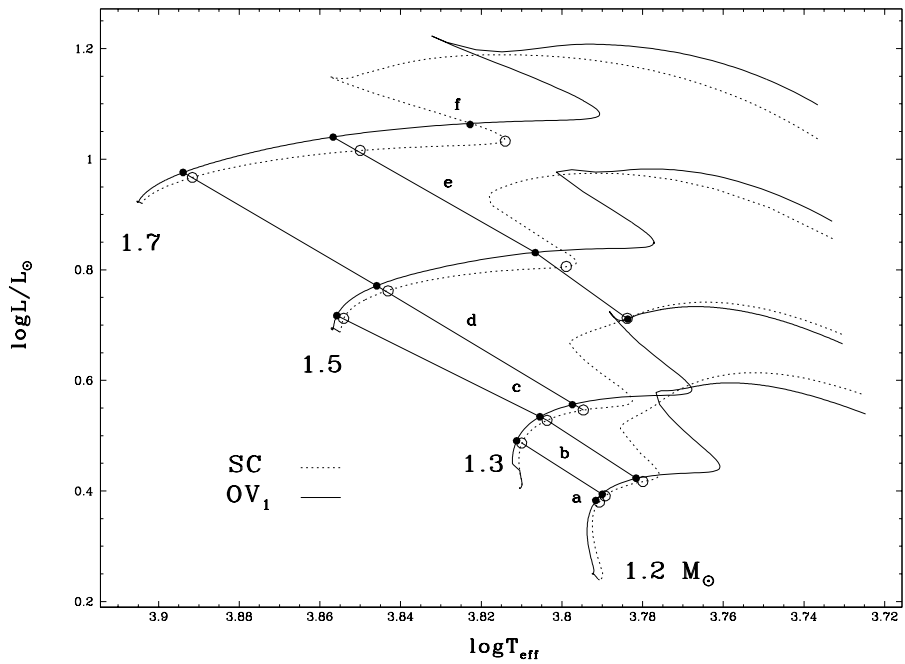,angle=-90,width=\linewidth}
\caption{Evolutionary tracks calculated with the two mixing scenarios, {\tt SC} and
{\tt OV$_1$}. The models selected for the survey of small separations are
marked with open and full circles for {\tt SC} and {\tt OV$_1$}, respectively.
The models of the same $\overline{D}_0$ are connected. The connecting lines are accompanied by
a letter, which indicates the value of mean large frequency separation, $\overline{D}_0.$
}
\label{f:3}
\end{figure}

In order to see how the different mixing scenarios influence the small separation, we consider
3D family of models in the mass range from $1.2$ to $1.7$ M$_\odot,$ which encompasses
most of the stars with expanding convective cores.\footnote{We exclude the stars with
higher masses, for which the expansion of convective core is negligible.} We assume the mixing
length parameter $\alpha=1.5$ and standard population I chemical composition (X=$0.7,$ Z=$0.02$).
The models are characterized by mass, the mean large frequency separation and the mixing scenario.
The grid of models considered in various mixing scenarios is given in Table~\ref{t:1}.
The models cover the early phase of the main sequence evolution which in the case of the
{\tt LY} and {\tt SC} scenarios is characterized by the convective core expansion.
Evolutionary tracks for selected masses and two mixing scenarios ({\tt SC, OV$_1$})
are shown in Figure~\ref{f:3}. There, the positions of the selected models are shown
with symbols. The tracks calculated with {\tt LY} scenario are nearly identical to
the tracks calculated with {\tt SC} scenario. The departure of the {\tt OV$_2$} tracks from
{\tt SC} tracks is similar to that of {\tt OV$_1$} tracks, but larger. The parameters
characterizing the evolutionary state of the core in selected models are given in
Table~\ref{t:2}.

\begin{table}
\centering
\begin{tabular}{|c|cccc|}
 \hline
  & \multicolumn{4}{|c|}{Mass [M$_\odot$]}  \\
 \hline
 $\overline{D}_0$ [$\mu$Hz] & 1.2 & 1.3 & 1.5 & 1.7 \\
 \hline
 93.8 & {\tt M2a} &           &           &           \\
 91.0 & {\tt M2b} & {\tt M3b} &           &           \\
 81.6 & {\tt M2c} & {\tt M3c} & {\tt M5c} &           \\
 74.5 &           & {\tt M3d} & {\tt M5d} & {\tt M7d} \\
 53.6 &           & {\tt M3e} & {\tt M5e} & {\tt M7e} \\
 42.3 &           &           &           & {\tt M7f} \\
 \hline
\end{tabular}
\caption{Grid and nomenclature of models used in the survey. The names of the models contain
decimal digit of the mass and a letter, which stands for the mean large frequency separation.}
\label{t:1}
\end{table}

\begin{sidewaystable}
\centering
\begin{tabular}{|c|cccc|cccc|cccc|cccc|}
 \hline
  & \multicolumn{4}{|c|}{\tt LY} & \multicolumn{4}{|c|}{\tt SC} &
  \multicolumn{4}{|c|}{\tt OV$_1$} & \multicolumn{4}{|c|}{\tt OV$_2$} \\
 \hline
  & $X_c$ & $x_{cc}$ & $m_{cc}$ &  &
    $X_c$ & $x_{cc}$ & $m_{cc}$ &  &
    $X_c$ & $x_{cc}$ & $m_{cc}$ &  &
    $X_c$ & $x_{cc}$ & $m_{cc}$ &  \\
 \hline
 {\tt M2a} & 0.291 & 0.046 & 0.021 & exp &  0.304 & 0.048 & 0.022 & exp &  0.425 & 0.088 & 0.104 & cont &  0.509 & 0.125 & 0.222 & cont \\
 {\tt M2b} & 0.248 & 0.048 & 0.024 & exp &  0.276 & 0.051 & 0.029 & exp &  0.390 & 0.085 & 0.101 & cont &  0.488 & 0.122 & 0.220 & cont\\
 {\tt M2c} & 0.083 & 0.047 & 0.036 & cont &  0.143 & 0.054 & 0.047 & cont &  0.266 & 0.077 & 0.100 & cont &  0.406 & 0.111 & 0.216 & cont\\\hline
 {\tt M3b} & 0.474 & 0.085 & 0.034 & exp &  0.479 & 0.086 & 0.036 & exp &  0.544 & 0.098 & 0.117 & cont &  0.587 & 0.132 & 0.248 & cont \\
 {\tt M3c} & 0.323 & 0.095 & 0.053 & exp &  0.335 & 0.097 & 0.055 & exp &  0.421 & 0.088 & 0.116 & cont &  0.504 & 0.120 & 0.230 & cont \\
 {\tt M3d} & 0.206 & 0.096 & 0.059 & cont &  0.223 & 0.099 & 0.052 & cont &  0.324 & 0.081 & 0.115 & cont &  0.440 & 0.111 & 0.222 & cont \\
 {\tt M3e} & 0.000 & 0.000 & 0.000 & -- &  0.000 & 0.000 & 0.000 & -- &  0.000 & 0.000 & 0.000 & -- &  0.185 & 0.082 & 0.197 & cont \\\hline
 {\tt M5c} & 0.565 & 0.083 & 0.079 & exp &  0.565 & 0.084 & 0.080 & exp &  0.592 & 0.108 & 0.154 & cont &  0.616 & 0.137 & 0.264 & cont \\
 {\tt M5d} & 0.466 & 0.079 & 0.085 & exp &  0.467 & 0.079 & 0.087 & exp &  0.510 & 0.100 & 0.151 & cont &  0.530 & 0.108 & 0.183 & cont \\
 {\tt M5e} & 0.065 & 0.050 & 0.064 & cont &  0.091 & 0.052 & 0.070 & cont &  0.237 & 0.074 & 0.137 & cont &  0.367 & 0.097 & 0.235 & cont \\\hline
 {\tt M7d} & 0.539 & 0.095 & 0.111 & exp &  0.546 & 0.096 & 0.112 & exp &  0.572 & 0.116 & 0.180 & cont &  0.592 & 0.140 & 0.278 & cont \\
 {\tt M7e} & 0.260 & 0.146 & 0.095 & cont &  0.270 & 0.069 & 0.096 & cont &  0.352 & 0.088 & 0.165 & cont &  0.417 & 0.107 & 0.249 & cont \\
 {\tt M7f} & 0.035 & 0.114 & 0.065 & cont &  0.054 & 0.048 & 0.069 & cont &  0.207 & 0.082 & 0.158 & cont &  0.316 & 0.089 & 0.238 & cont \\
 \hline
\end{tabular}
\caption{Characteristics of the convective core for the models of our survey. $X_c$ denotes
central hydrogen abundance, $x_{cc}$ -- fractional radius of the core, $m_{cc}$ - fractional
mass. "exp" and "cont" stands respectively for the core expanding and contracting (in mass).
For the {\tt OV} models, for which the chemical profile is smooth, we use the effective
size of the core.}
\label{t:2}
\end{sidewaystable}

\begin{figure}
\centering
   \epsfig{figure=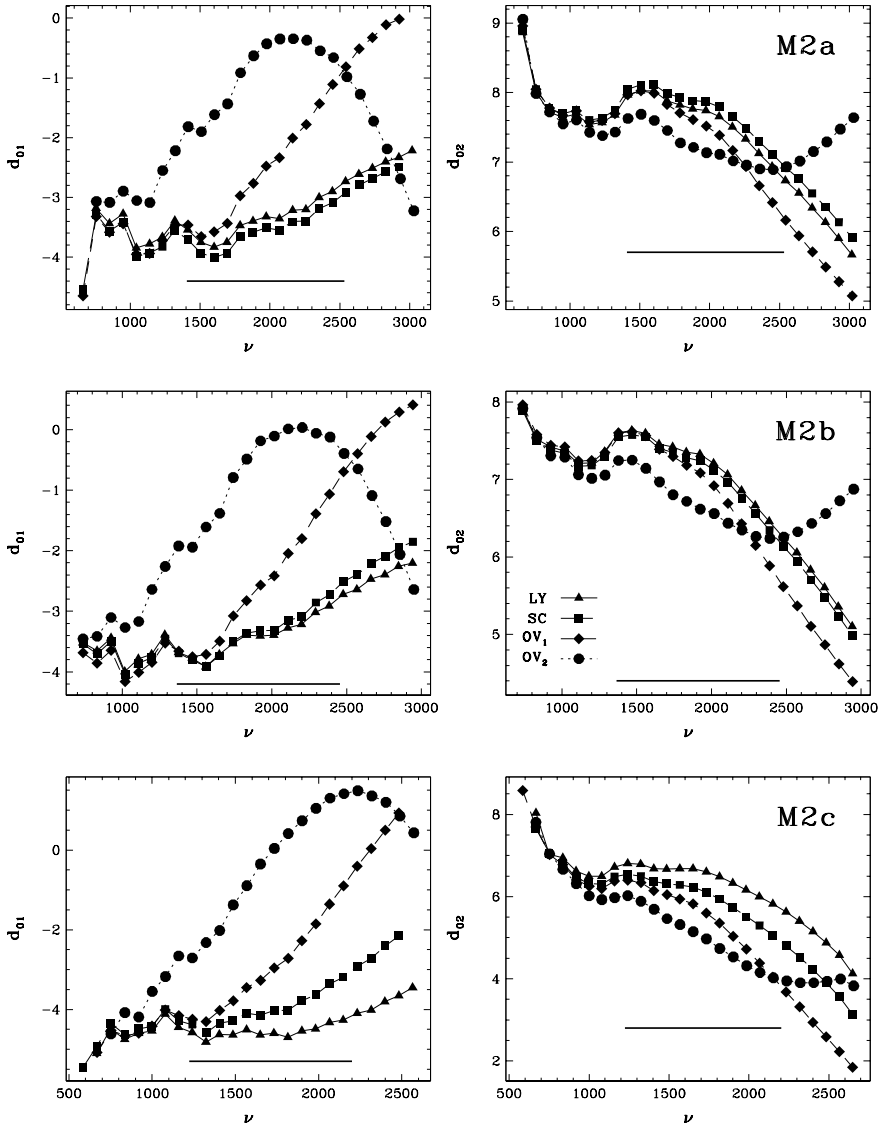,width=\linewidth}
\caption{Small separations for 1.2 M$_\odot$ models of the survey.
The rows correspond to models and the columns to different small separations,
$d_{01}$ and $d_{02}$.
Horizontal line on each panel spans the radial order range, $n=15\div 27$
-- see Section~\ref{s:41}}
\label{f:4}
\end{figure}

\begin{figure}
\centering
   \epsfig{figure=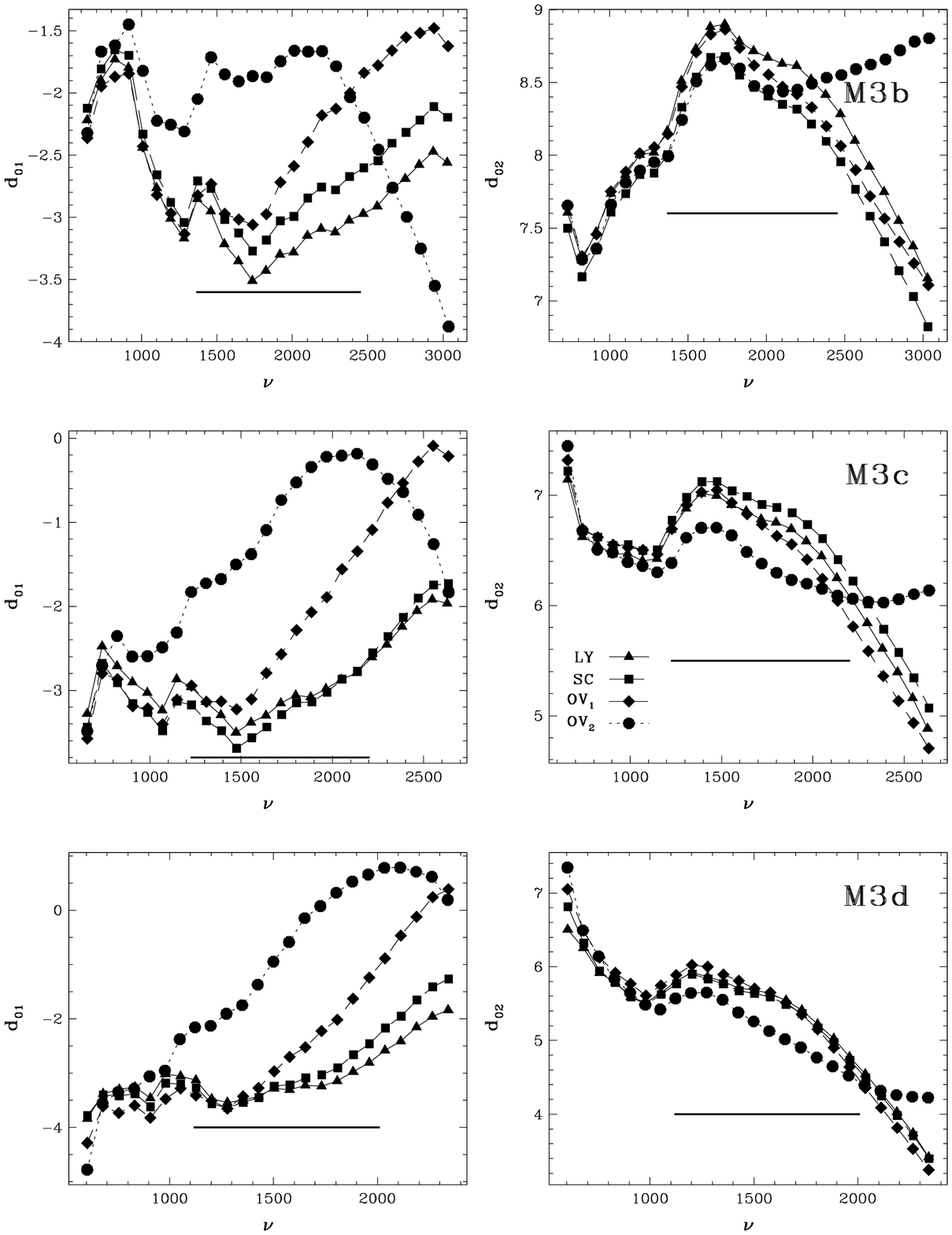,width=\linewidth}
\caption{Small separations for 1.3 M$_\odot,$ as in Figure~\ref{f:4}.}
\label{f:5}
\end{figure}

\begin{figure}
\centering
   \epsfig{figure=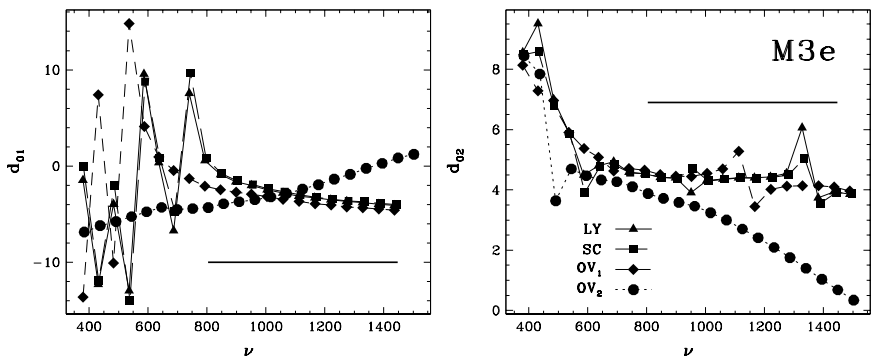,width=\linewidth}
\caption{Small separations for an evolved model of 1.3 M$_\odot,$ as in Figure~\ref{f:4}.}
\label{f:6}
\end{figure}

\begin{figure}
\centering
   \epsfig{figure=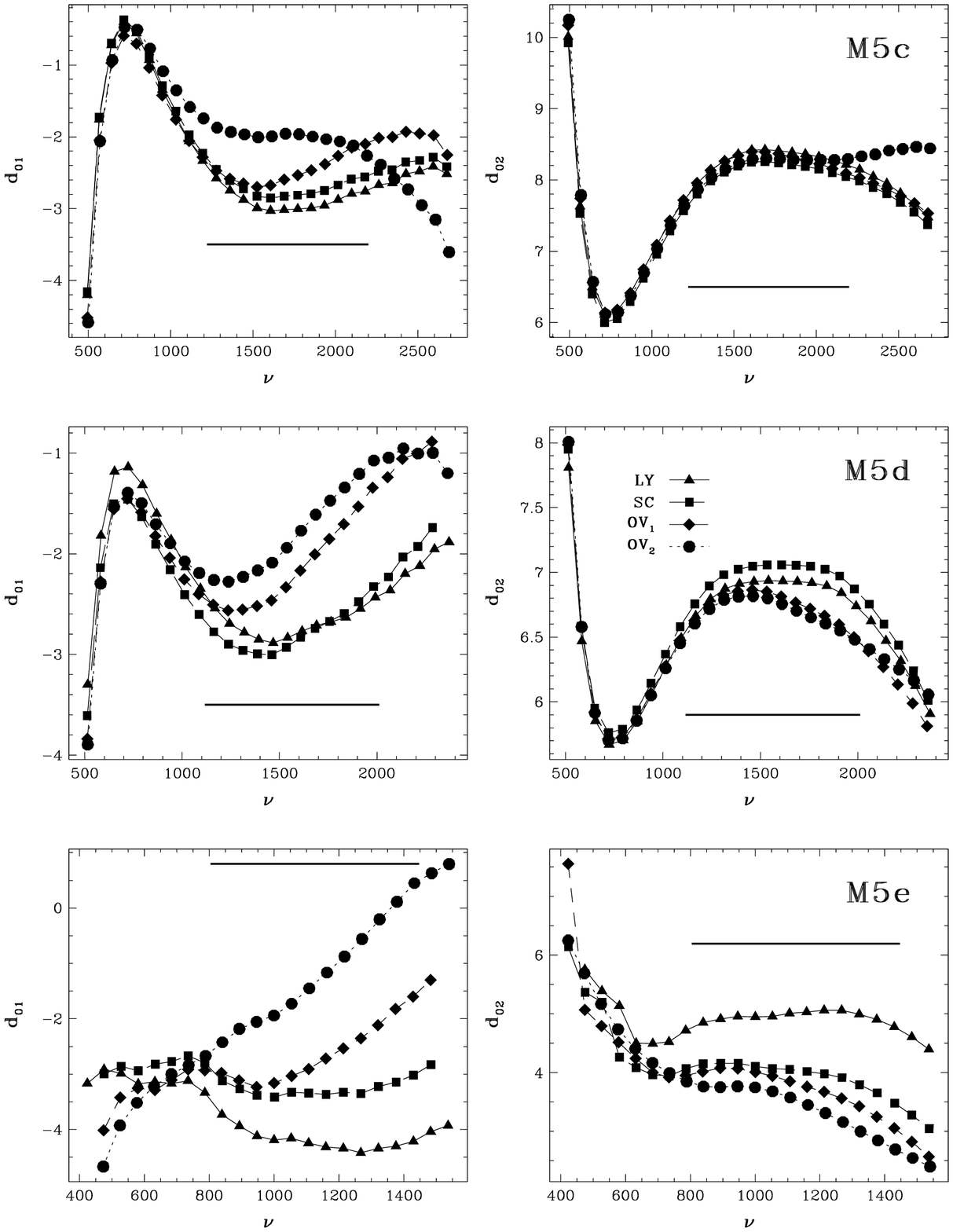,width=\linewidth}
\caption{Small separations for 1.5 M$_\odot,$ as in Figure~\ref{f:4}.}
\label{f:7}
\end{figure}

\begin{figure}
\centering
   \epsfig{figure=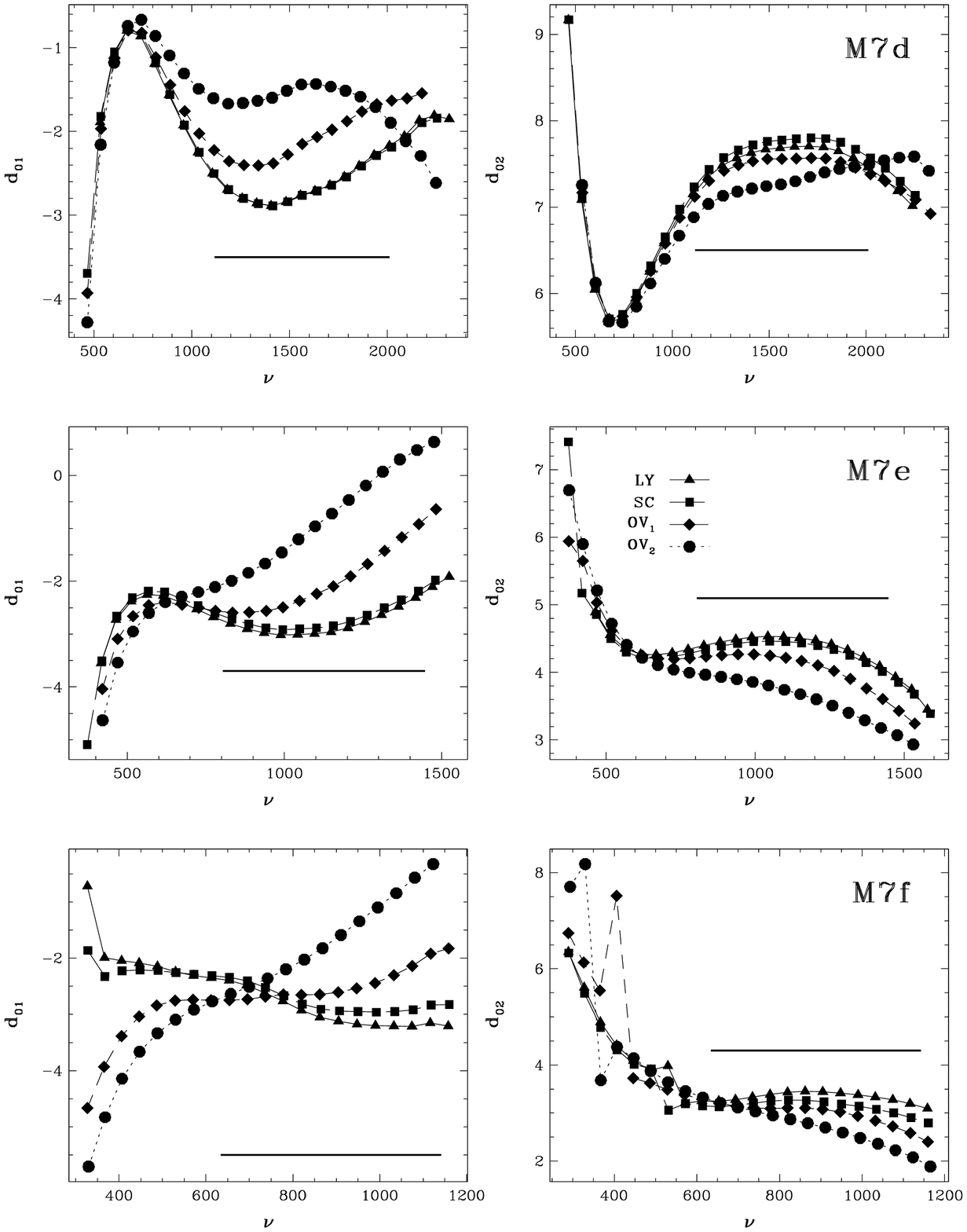,width=\linewidth}
\caption{Small separations for 1.7 M$_\odot,$ as in Figure~\ref{f:4}.}
\label{f:8}
\end{figure}

The $d(\nu)$ dependence in models of our survey are plotted in Figures~\ref{f:4} -- \ref{f:8}.
We may see that within the n-range there is a large variety in shapes of the $\nu$-dependence.
We see both monotonic increase and decrease, as well as occurrence of extrema.
There is a striking difference in the behavior of $d_{01}$ and $d_{02}.$ The former exhibits
significantly larger sensitivity for the adopted mixing scenario.
Therefore we concentrate our attention on the behavior of the $d_{01}$ separation.

\begin{figure}
\centering
   \epsfig{figure=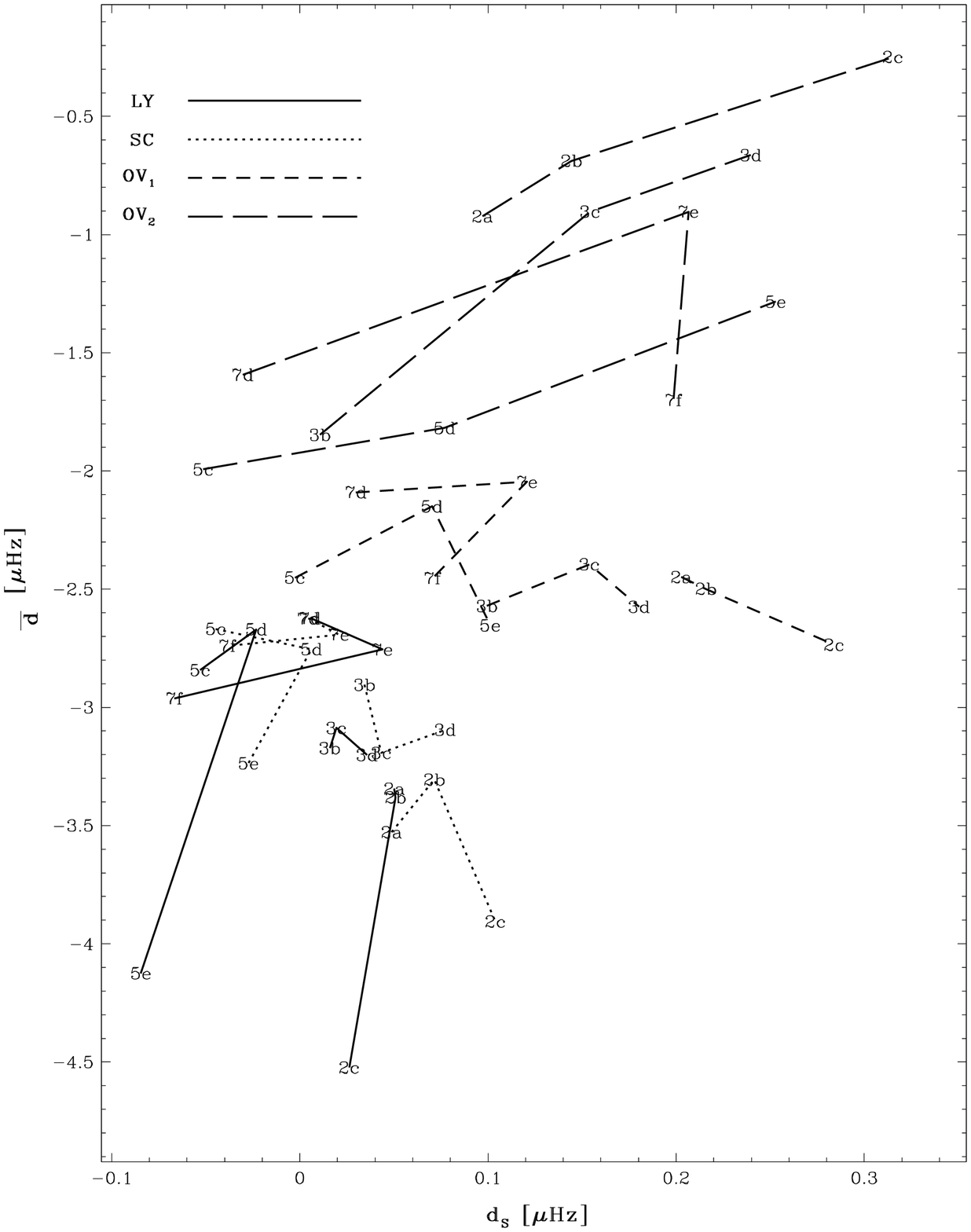,width=\linewidth}
\caption{Two parameters characterizing $d_{01}$ in models with various mixing
scenarios.
}
\label{f:9}
\end{figure}

We will characterize $d_{01}$ in various models by two parameters, the mean value over the
considered range of mode orders, $\overline{d},$ and the normalized mean slope,
$$d_S = \overline{D}_0 \dfrac{\partial d_{01}}{\partial \nu},$$
where the derivative is determined by the linear fit of $d_{01} (\nu).$
Now, each model is represented by
two quantities $\overline{d}$ and $d_S,$ which may be measured.
These are plotted in Figure~\ref{f:9} for all the models except of {\tt M3e},
which represents the post main-sequence phase.
The parameters $\alpha,$ X and Z influence the $\overline{d}-d_S$ diagram, but within the
reasonable ranges the effect is not large.
The most interesting feature is that models are grouped in the diagram according to the mixing
scenario. This means that even without precise knowledge of star parameters there
is a chance of discriminating between various scenarios. We see that the most negative
values of $\overline{d}$ correspond to the {\tt LY}
scenario, that is least mixed models. Moving upward we go in the direction of
more extensive mixing. The value of $d_S$ reflects primarily the evolutionary status
of the star.

Admittedly, the most difficult may be discriminating between the {\tt LY} and {\tt SC} scenarios.
We may see in Figure~\ref{f:9} that the largest differences in $\overline{d}$ between the models
calculated with the {\tt LY} and {\tt SC} scenarios at fixed mass and $\overline{D}_0$ do not
exceed $1\mu$Hz. There are various assessments of the accuracy of frequency measurements
in planned space missions. The optimistic error estimate of individual frequency data is
$0.1\mu$Hz (Baglin~\etal 2002). Thus there is a prospect for disentangling the two scenarios
but the objects must be carefully chosen. The models leading to the largest differences in
$\overline{d}$ are {\tt M2c} and {\tt M5e}, which correspond to the phase of vanishing
convective core. All models with intermediate masses between $1.2$ and $1.7$ M$_\odot$ in
the similar evolutionary phase show similar differences in $\overline{d}.$
None of the models with M$=1.3$ M$_\odot$ shown in Figure~\ref{f:9} reached this phase yet.
Thus the most suitable objects are stars near the end of the main-sequence phase.

Models calculated assuming extensive overshooting lay well above those calculated with either
{\tt LY} or {\tt SC} scenario. Thus there should be no difficulty in discriminating between
the extensive and weak mixing scenarios.

\begin{figure}
\centering
   \epsfig{figure=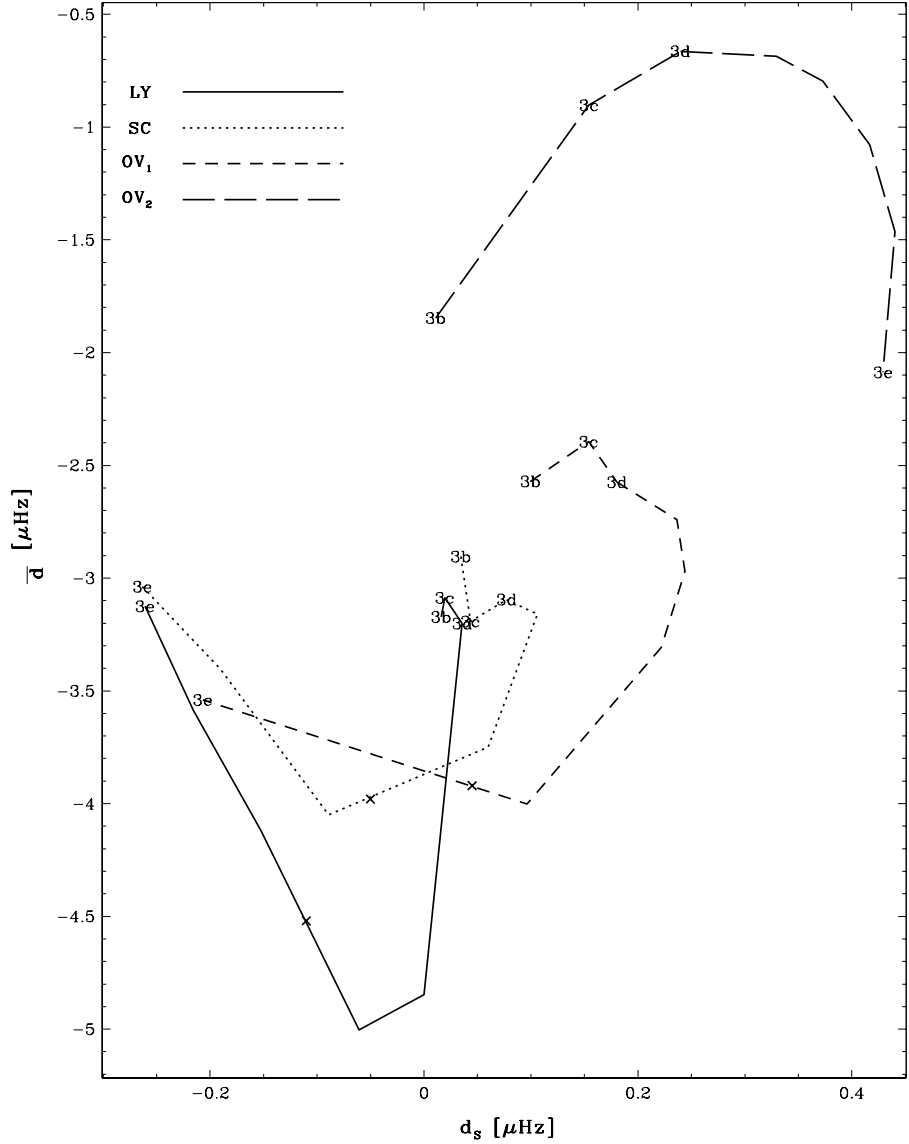,width=\linewidth}
\caption{Diagnostic diagram for $1.3$ M$_\odot.$
The same as in Figure~\ref{f:9} but including the most 
evolved models of our survey, ie {\tt M3e}. Cross indicates the TAMS model on each track
extending to the post main-sequence phase, ie {\tt LY}, {\tt SC} and
{\tt OV$_1$} tracks. All {\tt M3e} models, which are post main-sequence
stars, are located in the domain of low slopes.}
\label{f:10}
\end{figure}

The diagnostic utility of the $\overline{d}-d_S$ diagram is not restricted to main-sequence
stars. The signature of the mixing processes above an expanding core is present until the
hydrogen burning shell wipes it out. For example, the discontinuity arising in the {\tt LY}
scenario persists in the $1.3$ M$_\odot$ model for $0.6$ Gyr past the main-sequence phase,
which is about 1/5 of the main-sequence lifetime. Once the shell source enters the part
of the profile that is unreachable for mixing processes connected to the convective core,
all the memory of different mixing scenarios is erased.
Figure~\ref{f:10} shows the evolution of $1.3$ M$_\odot$ models in the $\overline{d}-d_S$
diagram through the main-sequence and early post main-sequence phases.
The evolutionary sequences for {\tt LY}, {\tt SC} and {\tt OV$_1$} nearly coincide in
most evolved models, {\tt M3e}, because in these models the signature of mixing processes
has been nearly erased. The {\tt OV$_2$} is still far away because the chemically inhomogeneous
zone extends well above the source.

Applications of the $\overline{d}-d_S$ diagrams is limited to the cases where the modes
partially trapped do not appear in the frequency domain of solar-like oscillations. Signature
of such partial trapping are rapid changes of the small separations, such as seen
in Figure~\ref{f:6}. The rapid changes are in fact a very sensitive probe of mixing processes,
but make our two parameters, $\overline{d}$ and $d_S,$ not well defined. Our calculations for
1.3 M$_\odot$ indicate that the post main-sequence phase, where we may trust in
$\overline{d}$ and $d_S,$ lasts about $0.6$ Gyr for weak mixing scenarios.

\section{Summary}
\label{s:5}

We have analyzed the small separations of a representative set of evolutionary models of
moderate (1.2 -- 1.9 M$_\odot$) mass stars
with different scenarios of mixing beyond the convective core. Four mixing scenarios 
were taken into account, layered scenario ({\tt LY}), semiconvective scenario ({\tt SC}) and
two overshooting scenarios of different extent ({\tt OV$_1$}, {\tt OV$_2$}).
Our numerical results for two types of small separations,
$d_{01}$ and $d_{02},$ indicate that within the
frequency range of p-modes, $d_{01}$ is more sensitive to mixing scenarios than $d_{02}.$
Our simple parametrization of $d_{01}$ by the mean value, $\overline{d},$ and the 
slope, $d_S,$ enables us to compare models and observations on a single $\overline{d}-d_S$
diagram.

We have found that the location of a star on the diagram depends on its
evolutionary status and the mixing scenario. During the main-sequence phase stars evolve slowly
toward higher slopes, $d_S.$
During this phase, stars having different mixing scenarios may be distinguished with
$\overline{d},$ which is correlated with the extent of mixing. The higher is the extent
of mixing, the larger is $\overline{d}.$ This correlation allows to determine the extent
of mixing with an accuracy depending on the accuracy of frequency measurements.  
For example, $0.2 H_P$ difference in the mixing extent manifests as a difference of
approximately $0.7\mu$Hz in $\overline{d}.$ Such an accuracy of $\overline{d}$ requires
the frequency accuracy of $1.8\mu$Hz or better. Optimistic predictions of the frequency
accuracy for space missions give a much better value of $0.1\mu$Hz.
The discrimination between two scenarios of low mixing extent, {\tt SC} and {\tt LY},
is difficult for young main-sequence objects because their values of $(\overline{d}, d_S)$
overlap. We show, however, that the distinction is possible for the most evolved main-sequence
stars. Similar {\tt SC} and {\tt LY} stars in this phase may differ as much as $1\mu$Hz in
$\overline{d},$ what requires the frequency accuracy of $2.6\mu$Hz or better.
Such stars are characterized by a rapid evolution of their cores. During this phase
such star moves rapidly towards low slopes on the $\overline{d}-d_S$ diagram.
Identification of the mixing scenario is still possible for such star but requires strong
constraints on other stellar observables like effective temperature, radius, metallicity and age.

\Acknow{This work was supported in part by the Polish State Committee for Scientific
Research grant 2-P03D-14.}


\end{document}